\documentclass{article}

\usepackage{microtype}
\usepackage{graphicx}
\usepackage{subcaption}
\usepackage{booktabs} 

\usepackage{hyperref}

\usepackage[preprint]{icml2026}

\usepackage{amsmath,amsfonts,bm}

\def\eqref#1{equation~\ref{#1}}

\def\1{\bm{1}}

\DeclareMathAlphabet{\mathsfit}{\encodingdefault}{\sfdefault}{m}{sl}
\SetMathAlphabet{\mathsfit}{bold}{\encodingdefault}{\sfdefault}{bx}{n}

\usepackage[utf8]{inputenc} 
\usepackage[T1]{fontenc}    
\usepackage{url}            
\usepackage{booktabs}       
\usepackage{amsfonts}       
\usepackage{nicefrac}       
\usepackage{microtype}      
\usepackage{xcolor}         
\usepackage{amsthm}

\newtheorem{theorem}{Theorem}[section]
\newtheorem{proposition}[theorem]{Proposition}

\usepackage{graphicx}
\usepackage{multirow}
\usepackage{xspace}
\usepackage{enumitem}
\usepackage[many]{tcolorbox}
\usepackage{fancyhdr}
\usepackage{kbordermatrix}
\usepackage{bm}
\usepackage[customcolors]{hf-tikz}
\usepackage{tikz}
\usetikzlibrary{shapes, arrows.meta, positioning, fit, backgrounds, calc,shadows,matrix}

\usepackage{textcomp}
\usepackage{wrapfig}
\usepackage{newfloat}
\usepackage{listings}
\newcommand{\eos}{\texttt{<EOS>}\xspace}
\newcommand{\NAME}{$\mathtt{BitHydra}$\xspace}   

\usepackage{color}
\usepackage{xcolor}
\def\red#1{\textcolor{red}{#1}}

\def\ie{$i.e.$}
\def\eg{$e.g.$}

\usepackage{hyperref}

 \hypersetup{
   colorlinks   = true,    
   urlcolor     = blue,    
   linkcolor    = blue,    
   citecolor    = blue      
 }

\icmltitlerunning{\NAME: Towards Bit-flip Inference Cost Attack against Large Language Models}
\begin{document}
\twocolumn[
  \icmltitle{\NAME: Towards Bit-flip Inference Cost Attack \\ against Large Language Models}
  \icmlsetsymbol{equal}{*}
    \begin{icmlauthorlist}
  \icmlauthor{Xiaobei Yan}{yyy}
  \icmlauthor{Yiming Li}{yyy}
  \icmlauthor{Hao Wang}{sch}
  \icmlauthor{Han Qiu}{sch}
  \icmlauthor{Tianwei Zhang}{yyy}
\end{icmlauthorlist}

\icmlaffiliation{yyy}{Nanyang Technological University \\ \texttt{xiaobei002@e.ntu.edu.sg, tianwei.zhang@ntu.edu.sg}}
\icmlaffiliation{sch}{Tsinghua University \\ \texttt{qiuhan@tsinghua.edu.cn}}

\icmlcorrespondingauthor{Yiming Li}{liyiming.tech@gmail.com}
   \vskip 0.3in
  ]
\printAffiliationsAndNotice{} 
\begin{abstract}
Large language models (LLMs) are widely deployed, but their substantial compute demands make them vulnerable to inference cost attacks that aim to deliberately maximize the output length. In this work, we investigate a distinct attack surface: maximizing inference cost by tampering with the model parameters instead of inputs. This approach leverages the established capability of Bit-Flip Attacks (BFAs) to persistently alter model behavior via minute weight perturbations, effectively decoupling the attack from specific input queries. To realize this, we propose \NAME, a framework that addresses the unique optimization challenge of identifying the exact weight bits that maximize generation cost. We formulate the attack as a constrained Binary Integer Programming (BIP) problem designed to systematically suppress the end-of-sequence (\ie, \eos) probability. To overcome the intractability of the discrete search space, we relax the problem into a continuous optimization task and solve it via the Alternating Direction Method of Multipliers (ADMM). We evaluate \NAME across 10 LLMs (1.5B–16B). Our results demonstrate that the proposed optimization method efficiently achieves endless generation with as few as 1-4 bit flips on all testing models, verifying the effectiveness of the ADMM-based formulation against both standard models and potential defenses.

\end{abstract}

\section{Introduction}
\label{sec:Introduction}

Large Language Models (LLMs)~\cite{carlini2021extracting,ouyang2022training,touvron2023llama} have demonstrated remarkable capabilities across a wide range of real-world applications, including online chat~\cite{shen2023chatgpt}, customer service~\cite{gimpel2023unlocking}, and financial services~\cite{wu2023bloomberggpt}. As LLMs are increasingly deployed through cloud-based ML-as-a-Service (MLaaS) platforms, minimizing inference cost has become critical for both service providers and end-users—enhancing service availability and reducing token-based billing costs.
However, previous studies have shown that deep neural networks are vulnerable to inference cost attacks~\cite{shumailov2021sponge,shapira2022denial,shapira2023phantom,liu2023slowlidar, schoof2024beyond, xiao2024sponge, ma2024slowtrack,muller2024impact}, where the attacker crafts malicious input to maximize the latency and cost of the victim model's inference execution. Such attacks can lead to substantial operational overhead for service providers and degrade the user experience. Recently, researchers designed inference cost attacks against auto-regressive LLMs~\cite{feng2024llmeffichecker,geiping2024coercing,dong2024engorgio,kumar2025overthink} and multimodal LLMs~\cite{gao2024inducing}. As the victim model's inference cost scales with the response length, the attacker's objective is to mislead the model to generate as many tokens as possible using short induced prompts. 

Despite their diversity, existing inference cost attacks share a key feature: they rely on specially-crafted inputs to induce excessive computation, which are inherently self-targeting: the attacker, who submits the adversarial prompt, will be charged for the long generated responses, bearing the inference cost. Also, to achieve damages to other users and service providers at scale, the attacker needs to consistently send a large volume of malicious input, which can be costly and easy to spot. 
To overcome this, we propose the \textbf{Bit-Flip Inference Cost Attack (BICA)}, a new paradigm of inference cost attacks that target the model parameters directly. 
Our approach leverages the mature mechanism of Bit-Flip Attacks (BFAs)~\cite{rakin2019bit,yao2020deephammer,dong2023one}—a realistic hardware-level threat, capable of inducing severe consequences (\eg, accuracy degradation or targeted misclassification) by flipping a few critical bits in memory. Our BICA repurposes this proven attack vector with a new strategy for a novel objective: rather than degrading model correctness, we optimize bit-flips to maximize computational latency. It aims to modify just a few weight bits to let the model persistently generate excessive output for \emph{all} users, decoupling the attacker from the inference cost. Because these manipulations occur at the physical hardware level, they bypass conventional software-based monitoring, making BICA both stealthy and persistent.

However, implementing such BICAs introduces several technical challenges, including: \textbf{(1) Effectiveness}: how to design an effective loss function that encourages LLMs to generate substantially longer outputs; \textbf{(2) Optimization Hardness}: how to identify the optimal set of critical bits from the vast parameter space of LLMs, which is inherently a discrete combinatorial problem; \textbf{(3) Fidelity}: how to ensure that, even after flipping these critical bits, the victim model continues to produce outputs that appear benign and exhibit no obvious anomalies; and \textbf{(4) Efficiency}: how to minimize the search overhead required to locate these vulnerabilities within models containing billions of parameters.

To tackle these challenges, we propose \NAME, a rigorous optimization-based attack framework. To ensure high Effectiveness, we introduce a specialized loss function, $\mathcal{L}_{\eos}$, designed to suppress the probability of the end-of-sequence (\eos) token. Minimizing this loss forces the victim LLM to delay termination, producing abnormally long outputs without significantly degrading the coherence of the generation.
To resolve the scalability and combinatorial complexity (Optimization Hardness), \NAME departs from heuristic greedy methods. We instead formulate the critical bit identification as a constrained \textit{Binary Integer Programming (BIP)} problem, and solve it via the \textit{$\ell_p$-Box ADMM} framework~\cite{wu2018ell} which relaxes binary constraints into continuous intersection constraints. 
To guarantee efficiency, we restrict the optimization domain exclusively to the output embedding weights corresponding to the \eos token. This targeted scope drastically reduces the memory and computational resources required, avoiding the prohibitive cost of searching the entire parameter space.
Consequently, \NAME identifies the exact bits within the \eos embedding that maximize inference cost. Besides, the problem's explicit cardinality constraint restricts modifications to a minimal budget, thereby preserving the model's fidelity and ensuring the attack remains stealthy.

In summary, our main contributions are four-fold. 

\begin{itemize} [leftmargin=*, itemsep=2pt, topsep=2pt, parsep=0pt]

\item We revisit existing inference cost attacks and reveal their inherent limitations and underlying reasons. 

\item We propose a new inference cost attack paradigm, \ie, bit-flip inference cost attack (BICA), that targets model parameters rather than inputs, allowing large-scale persistent damages that affect all users. 

\item We design \NAME, which formulates the attack as a constrained BIP problem and solves it via an ADMM-based continuous relaxation to effectively suppress the \eos token with a minimal bit-flip budget. 

\item We demonstrate the effectiveness of \NAME through extensive experiments, showing that it causes 100\% of evaluation prompts to reach the maximum generation length on representative LLMs like Llama3-8B while requiring only \textit{one} bit flip. We also demonstrate \NAME's transferability to unseen prompts, suggesting a generalizable and systemic shift in generation dynamics.
\end{itemize}

\section{Background and Related Work}
We present the background of inference cost attacks and BFAs in this section. Additional information about LLM and its data representation can be found in Appendix~\ref{appendix:bgd}.

\label{sec:Background}

\subsection{Inference Cost Attacks}

Inference cost attacks aim to exploit the compute-intensive nature of deep learning models to intentionally increase the models' latency or resource consumption during inference, ultimately leading to high compute cost and degraded user experience. \citet{shumailov2021sponge} introduced the concept of \textit{sponge examples} and designed the first inference cost attack.
Later works extended this attack across various tasks and domains, such as image understanding~\cite{chen2022nicgslowdown}, object detection~\cite{
xiao2024sponge, ma2024slowtrack}, and language translation~\cite{chen2022nmtsloth}.

Recent studies showed that this inference-cost threat is amplified in LLMs. For example, LLMEffiChecker~\cite{feng2024llmeffichecker} employed gradient-guided search to find minimal, imperceptible input perturbations that raise inference cost; \citet{geiping2024coercing} coerced LLMs into generating specific starting responses, indirectly imposing higher computational cost; \citet{dong2024engorgio} designed adversarial prompts that prolong decoding in modern autoregressive LLMs; \citet{gao2024inducing} crafted verbose images that elevate latency and energy use in multimodal LLMs; and \citet{kumar2025overthink} intentionally induced model `overthinking', slowing its reasoning process.

To our best knowledge, all existing attacks induce damage by manipulating the model inputs, which leads to two practical limitations. (1) Modern LLM services use token-based billing; for example, OpenAI's o3 API charges \$10 per 1 million input tokens and \$40 per 1 million output tokens~\cite{chatgptpricing}. Thus, while abnormally long outputs increase provider's computational load, attacker ultimately pays the bill, and provider suffers only mild externalities. (2) Each adversarial input affects its own inference, offering no persistent, cross-user impact. These limitations substantially reduce the practical severity of such attacks.

\vspace{-8pt}

\subsection{Bit-Flip Attacks via Rowhammer}
Bit-flip attacks (BFAs) are hardware-level attacks that tamper with critical bits in DRAM. A prominent vector is Rowhammer~\cite{kim2014flipping}, which rapidly activates aggressor rows to disturb adjacent cells and flip bits, even in the presence of common error-correction schemes~\cite{gruss2018another,cojocar2019exploiting}. In particular, such faults can be triggered \textit{without} physical access to the device in practice, by running malicious code that repeatedly hammers memory on commodity CPUs~\cite{
jattke2022blacksmith,half-double} and GPUs with GDDR5~\cite{lin2025gpuhammer} or HBM~\cite{HBM2_RH_Profile}

In the context of machine learning, attackers apply BFAs to flip selected bits in the parameters of a deployed model. Existing attacks are commonly categorized by the objectives: untargeted attacks \cite{rakin2019bit,chen2023unveiling,li2024yes} degrade overall model performance, whereas targeted attacks \cite{
bai2021targeted,dong2023one,coalson2024prisonbreak} steer a model's behavior in specific ways, such as forcing misclassification or overriding content filters. To achieve precise and effective bit flips, attackers commonly pair Rowhammer with system-level memory placement tricks that rely on legitimate operating system features, \eg, leveraging the page cache~\cite{li2024yes}, memory deduplication~\cite{razavi2016flip}, or per-CPU page-frame caches~\cite{rakin2022deepsteal}. For instance, by first ensuring the model weights are resident in DRAM via the page cache and then inducing flips in those pages, attackers corrupt the in-memory copy so that subsequent loads by the victim process transparently retrieve the tampered weights from memory rather than the pristine file on disk.

Despite substantial progress on DNNs, and their feasibility for LLMs remains largely underexplored. More importantly, prior work targets accuracy degradation or specific misbehavior, which differs fundamentally from our objective: inflating the computational cost of ordinary queries while preserving task correctness. Existing BFA techniques are ill-suited for this purpose as they do not identify weight bits that modulate execution paths or computational complexity. Designing an effective bit-flip inference-cost attack for LLMs remains an important open problem.

\vspace{-1em}
\section{Problem Formulation and Analysis}
In our paradigm, an adversary flips a small set of cost-critical bits in a target LLM, biasing it to produce longer responses to \emph{any} prompt and thereby scaling provider-side computation. We term this the bit-flip inference-cost attack (BICA). Below we formalize the threat model and analyze the key challenges in realizing BICA.

\subsection{Threat Model}
\label{sec:Threat Model}

\noindent \textbf{Attacker's Goal}. The attacker aims to inflate the inference cost of a deployed LLM persistently and at scale, without compromising task accuracy. Specifically, the objective is to induce the model to generate abnormally long responses for \emph{any} user prompt, thereby amplifying computational overhead across users and sessions. 
Such an attack poses serious risks to both users and LLM-integrated service providers. For users, it leads to elevated query costs and significantly increased latency, particularly detrimental for time-sensitive applications, ultimately degrading the overall user experience. For providers, the attack escalates operational costs and may cause query congestion due to slower response time. Over time, reduced service availability and increased expenses may lead to user attrition and reputational harm.

\noindent \textbf{Attacker's Capacity}. We adopt the standard Bit-Flip Attack (BFA) threat model established in prior literature~\cite{rakin2019bit,yao2020deephammer,liu2023neuropots,li2024yes}. 
The adversary is hereby modeled as an unprivileged tenant co-located with the victim LLM on a shared physical server (\eg, in a public MLaaS environment). Consistent with well-documented hardware vulnerabilities, the attacker can induce precise bit flips in DRAM via Rowhammer-style fault injection~\cite{kim2014rowhammer} without requiring physical access or elevated system privileges. Following the standard assumption for bit-flip-based weight attacks, we assume that the attacker has access to model's architecture and weights.

\subsection{Main Challenges of Instantiating BICA}  
\label{sebsec:challenges}

Building a bit-flip inference cost attack imposes more strict constraints than traditional accuracy-degrading BFAs. A successful method must \textbf{(1)} inflate the output length under normal usage while \textbf{(2)} preserving functional plausibility \textbf{(3)} under a very small flip budget to remain practical and stealthy. In our early exploration, we attempted a brute-force strategy that scans the entire weight space for all potential bits and measures their effects. This naive approach revealed three fundamental obstacles: catastrophic numerical failures, visible degradation of linguistic quality, and prohibitive search cost, which collectively motivate a structure-aware design introduced by our method.

\noindent \textbf{Challenge 1 (Catastrophic Numerical Failures)}.
Flipping arbitrary bits frequently drives the LLM into catastrophic model states, with decoding collapsing to `\texttt{NaN}' after only a few flips in many cases. This failure mode is uncommon in traditional BFAs in attacking feedforward CNN/MLP settings but is amplified in LLMs due to their autoregressive nature and tightly coupled operations (\eg, LayerNorm, Softmax, attention scaling) over long sequences. A perturbed early-layer weight can be magnified through normalization and exponentiation, triggering overflow/underflow or near-zero variance divisions; the instability then recurs across decoding steps, culminating in the \texttt{NaN} outputs.

\noindent \textbf{Challenge 2 (Visible Degradation of Linguistic Quality)}.
Even when the model does not crash, bit-flipping often yields incoherent text, such as garbled symbols, broken tokens, and non-linguistic artifacts. This indicates that bit flips scattered throughout the model can disrupt semantic and syntactic alignment, degrading internal representations beyond recovery. Unlike vision models, where spatial redundancies/correlations can buffer mild corruption, LLMs lack comparable structural slack, so small weight modifications can visibly erode linguistic fidelity. A successful BICA must therefore identify critical bits that lengthen the output while preserving generation plausibility and task utility.

\noindent \textbf{Challenge 3 (Prohibitive Search Cost)}.
Exhaustively scanning and evaluating bits in large-scale LLMs with billions of parameters is computationally prohibitive. Loading full weight matrices for gradient- or search-based scoring, running per-flip impact tests, and measuring downstream cost inflation impose heavy memory and latency overheads. A viable BICA requires an efficient search strategy that narrows the candidate space and prioritizes \emph{cost-critical} locations, achieving persistent cost inflation with a small flip budget.

\section{Methodology}
\label{sec:Methodology}
\subsection{Overall Workflow}
\label{subsec:workflow}

The \NAME framework operates through an iterative pipeline that alternates between continuous relaxation and discrete modification. As illustrated in Figure~\ref{fig:workflow}, the entire process (comprising both the ADMM optimization (Phase 1) and the bit selection (Phase 2)) is repeated for each bit flip. The workflow proceeds as follows:
\begin{enumerate}[leftmargin=*, noitemsep, topsep=0pt]
    \item \textbf{Input:} Original quantized weight matrix $\bm{B}_{orig}$.
    \item \textbf{Phase 1 (Continuous Optimization):} We initialize the ADMM variables with the current weights. We then solve the relaxed continuous optimization problem to find the optimal search direction (\ie, $-\nabla \mathcal{L}$) and the continuous proxy weights $\hat{\bm{B}}$ that maximize the attack objective given the current state.
    \item \textbf{Phase 2 (Discrete Selection):} Leveraging the gradients and proxy values from Phase 1, we rank and filter candidate bits. The top candidates undergo empirical evaluation, where we apply them to the model and measure the resulting generation length. The single most effective bit is permanently flipped to produce $\bm{B}^{(t+1)}$. This updated weight matrix serves as the input for the next iteration, and the cycle repeats $k$ times until the flip budget is exhausted or the attack is sufficiently effective.
\end{enumerate}

\begin{figure}[t]
    \centering
    \resizebox{0.95\columnwidth}{!}{%
    \begin{tikzpicture}[
        font=\sffamily,
        >=Latex,
        node distance=0.5cm and 0.6cm, 
        step_box/.style={
            rectangle, 
            draw=black!70, 
            fill=white, 
            rounded corners=2pt, 
            align=center, 
            font=\tiny,
            drop shadow={opacity=0.1},
            inner sep=2pt,
            text width=1.3cm,
            minimum height=0.8cm
        },
        mat/.style={
            matrix of nodes, 
            nodes={
                draw=gray!30, 
                fill=white, 
                minimum size=4mm, 
                anchor=center, 
                font=\ttfamily\tiny, 
                inner sep=0pt
            }, 
            row sep=-\pgflinewidth, 
            column sep=-\pgflinewidth,
            ampersand replacement=\&
        },
        static/.style={
            text=gray!50,
            fill=gray!5
        },
        target/.style={
            text=black,
            fill=white,
            font=\ttfamily\tiny\bfseries
        },
        phase/.style={
            rounded corners=4pt, 
            inner sep=4pt,
            fill opacity=0.4
        },
        arr/.style={
            ->, 
            thick, 
            draw=black!70, 
            rounded corners=2pt
        }
    ]

    
    \node (lbl_start) [font=\bfseries\scriptsize] {Current Weights $\bm{B}^{(t)}$};
    \matrix (mat_start) [mat, below=0.05cm of lbl_start] {
        |[static]| 1 \& |[static]| 0 \& |[static]| 1 \& |[static]| 0 \\
        |[target]| 0 \& |[target]| 1 \& |[target]| 0 \& |[target]| 1 \\
        |[static]| 1 \& |[static]| 1 \& |[static]| 0 \& |[static]| 0 \\
    };
    \node [left=0.05cm of mat_start-2-1, font=\tiny, text=black, align=right] {\textbf{EOS Row}\\(Active)};

    \node (step_relax) [step_box, right=0.6cm of mat_start, anchor=west, yshift=0.6cm] {
        \textbf{1. Relax} \\[-0.1em]
        $\min \mathcal{L}_{\text{eos}} + \mathcal{L}_{\text{cons}}$
    };
    
    \node (step_solve) [step_box, below=0.3cm of step_relax] {
        \textbf{2. Solve} \\[-0.1em]
        $\hat{\bm{B}} \leftarrow \text{prox}_{\mathcal{L}}$ \\
        $\bm{U} \leftarrow \Pi_{\mathcal{S}}$
    };

    \matrix (mat_proxy) [mat, right=0.6cm of step_solve] {
        |[static]| .9 \& |[static]| .1 \& |[static]| .8 \& |[static]| .2 \\
        |[target]| .1 \& |[target, fill=blue!10, draw=blue!60, thick]| \textbf{.6} \& |[target]| .1 \& |[target]| .9 \\
        |[static]| .8 \& |[static]| .9 \& |[static]| .1 \& |[static]| .1 \\
    }; 
    \node (lbl_proxy) [above=0.05cm of mat_proxy, font=\bfseries\scriptsize] {Proxy $\hat{\bm{B}}$};

    \draw [arr] (mat_start.east) -- ++(0.3,0) |- (step_relax.west);
    \draw [arr] (step_relax.south) -- (step_solve.north);
    \draw [arr] (step_solve.east) -- (mat_proxy.west);

    \begin{scope}[on background layer]
        \node (bg_p1) [phase, fill=cyan!10, fit=(step_relax)(step_solve)(mat_proxy)(lbl_proxy), draw=cyan!20] {};
        \node [anchor=south, font=\bfseries\scriptsize, text=cyan!50!black, inner sep=1pt] at (bg_p1.north) {Phase 1: Continuous Optimization};
    \end{scope}


    \matrix (mat_diff) [mat, below=0.6cm of mat_proxy] { 
        |[static]| - \& |[static]| - \& |[static]| - \& |[static]| - \\
        |[target]| .1 \& |[target, fill=orange!20, draw=orange!60, thick]| \textbf{.4} \& |[target]| .1 \& |[target]| .1 \\
        |[static]| - \& |[static]| - \& |[static]| - \& |[static]| - \\
    };
    \node (lbl_diff) [below=0.05cm of mat_diff, font=\bfseries\scriptsize] {Deviation $\Delta$};
    
    \node (step_score) [step_box, left=0.6cm of mat_diff] {
        \textbf{3. Score} \\[-0.1em]
        $\Delta \leftarrow |\hat{\bm{B}} - \bm{B}|$ \\
        $S \leftarrow \Delta \times \text{Sig}$
    };
    
    \node (step_select) [step_box, below=0.3cm of step_score] {
        \textbf{4. Select} \\[-0.1em]
        $i^* \leftarrow \arg\max(S)$
    };
    
    \path (step_score.south) -- (step_select.north) coordinate[midway] (center_stack);
    \matrix (mat_new) [mat] at (mat_start |- center_stack) { 
        |[static]| 1 \& |[static]| 0 \& |[static]| 1 \& |[static]| 0 \\
        |[target]| 0 \& |[target, fill=green!20, draw=green!60!black, thick]| \textbf{0} \& |[target]| 0 \& |[target]| 1 \\
        |[static]| 1 \& |[static]| 1 \& |[static]| 0 \& |[static]| 0 \\
    };
    \node (lbl_new) [above=0.05cm of mat_new, font=\bfseries\scriptsize] {Updated $\bm{B}^{(t+1)}$};
    \node [below=0.05cm of mat_new, font=\tiny, text=green!50!black] {Bit Flipped ($1 \to 0$)};

    
\draw [arr] ([xshift=-0.3cm]mat_proxy.south east) -- ([xshift=-0.3cm]mat_diff.north east);
    
    \draw [arr] (mat_diff.west) -- (step_score.east);
    \draw [arr] (step_score.south) -- (step_select.north);
    
    \draw [arr] (step_select.west) -- ++(-0.3,0) |- (mat_new.east) node[pos=0.75, above, font=\tiny] {Flip};

    \begin{scope}[on background layer]
        \node (bg_p2) [phase, fill=orange!10, fit=(mat_diff)(step_score)(step_select)(lbl_diff), draw=orange!20] {};
        \node [anchor=south, font=\bfseries\scriptsize, text=orange!50!black, inner sep=1pt] at (bg_p2.north) {Phase 2: Discrete Selection};
    \end{scope}

    \draw [arr, dashed, draw=purple!60, thick] (mat_new.west) -- ++(-0.4, 0) |- ([yshift=-0.3cm]mat_start.west);
    \node [text=purple!80, font=\bfseries\tiny, rotate=90, anchor=south] at ($(mat_new.west)+(-0.4, 1.5)$) {Iterate until budget $k$};

    \end{tikzpicture}
    }
    \caption{The iterative workflow of \NAME. \textbf{Left Matrices} represent weights of the output embedding (only the \textit{EOS Row} is updated). \textbf{Phase 1} relaxes and solves the problem via ADMM updates (Steps 1 \& 2). \textbf{Phase 2} computes the deviation matrix ($\Delta$) by comparing the continuous proxy $\hat{\bm{B}}$ with the current binary weights. Candidate bits are then scored and selected for flipping.}
    \vspace{-10pt}
    \label{fig:workflow}
\end{figure}

\subsection{Detailed Mechanisms}

\noindent \textbf{Attack Objective.}
To encourage prolonged generation, we hereby introduce a loss function $\mathcal{L}_{\eos}$ that penalizes the probability of output termination. We specifically target the normalized likelihood of the end-of-sequence (\eos) token, summed over the decoding sequence:
\begin{equation}
\label{eq:eosloss}
\mathcal{L}_{\eos}(\hat{\bm{B}}) = \sum_{i=1}^{N} \text{Softmax}(f_i^{\eos}(\bm{x}; \hat{\bm{B}})),
\end{equation}
where $\bm{x}$ is the input context, $N$ is the sequence length, and $f_i^{\eos}(\cdot)$ is the logit assigned to \eos at step $i$ given the perturbed weights $\hat{\bm{B}}$. We utilize the normalized Softmax probability rather than raw logits to effectively capture the relative likelihood of termination against competing tokens.

To minimize semantic damage and computational overhead, we restrict our modifications solely to the output embedding row corresponding to the \eos token, denoted as $\bm{W}_o[\eos] \in \mathbb{R}^d$. This targeted approach ensures functional stealthiness by preserving the relative ranking of all other tokens.
Consider the perturbed logit vector $\bm{l}'$ resulted from a modification $\Delta \bm{W}$ to the \eos row:
\begin{equation}
\label{eq:logit_modified}
l'(i) = 
\begin{cases}
(\bm{W}_o[\eos] + \Delta \bm{W}) \cdot \bm{h}, & \text{if } i = \eos \\
\bm{W}_o[i] \cdot \bm{h}, & \text{otherwise}
\end{cases},
\end{equation}
where $\bm{h}$ is the hidden state. Since the logits for all non-\eos tokens ($i \neq \eos$) remain unchanged, the ratio of their probabilities is preserved:
\begin{equation}
\label{eq:prob_ratio}
\frac{P(i)}{P(j)} = \frac{e^{l'(i)}}{e^{l'(j)}} = \frac{e^{l(i)}}{e^{l(j)}}, \quad \forall i,j \neq \eos.
\end{equation}
Consequently, the model's internal preference between any two benign tokens remains identical to the clean model. 
This guarantees that \NAME suppresses termination while minimizing disruption to the fluency or coherence of the generated content.

In quantized LLMs (\eg, \texttt{int8}), the \eos output embedding vector is stored as a binary matrix $\bm{B} \in \{0, 1\}^{d \times Q}$, where $d$ is the hidden dimension and $Q$ is the quantization bit-width. In this case, our objective is to find a perturbed binary configuration $\hat{\bm{B}}$ that minimizes $\mathcal{L}_{\eos}$ subject to a bit-flip budget $k$. Formally, we formulate this as a Binary Integer Programming (BIP) problem:
\begin{equation}
\label{eq:bip_obj}
\begin{split}
    \min_{\hat{\bm{B}}} \quad & \mathcal{L}_{\eos}(\hat{\bm{B}}) \\
    \text{s.t.} \quad & \hat{\bm{B}} \in \{0, 1\}^{d \times Q}, 
                     \|\hat{\bm{B}} - \bm{B}_{orig}\|_0 \le k,
\end{split}
\end{equation}
where $\bm{B}_{orig}$ represents the original clean weight bits, and $\|\cdot\|_0$ denotes the $\ell_0$-norm (Hamming distance). This formulation seeks the optimal set of at most $k$ bit flips that maximally suppresses the \eos probability.

\noindent \textbf{ADMM-based Relaxation.}
Solving Eq.~(\ref{eq:bip_obj}) directly is NP-hard due to the discrete binary and cardinality constraints. To address this, we exploit the $\ell_p$-Box ADMM framework~\cite{wu2018ell}, which relaxes the discrete constraints into continuous ones solvable via the Alternating Direction Method of Multipliers (ADMM). Specifically, we first replace the binary constraint $\hat{\bm{B}} \in \{0, 1\}^{d \times Q}$ with the intersection of two continuous sets: the box constraint $\mathcal{S}_b = [0, 1]^{d \times Q}$ and the $\ell_2$-sphere constraint $\mathcal{S}_p = \{ \bm{X} : \| \bm{X} - \mathbf{0.5} \|_2^2 = \frac{dQ}{4} \}$. 
Then we transform the inequality constraint on bit-flips (\ie,$\|\hat{\bm{B}} - \bm{B}_{orig}\|_0 \le k$) into an equality constraint by introducing a non-negative slack variable $u_3 \ge 0$. The equivalent problem is formulated as:
\begin{equation}
\label{eq:admm_formulation}
\begin{split}
    \min_{\hat{\bm{B}}, \bm{U}_1, \bm{U}_2, u_3} \quad & \mathcal{L}_{\eos}(\hat{\bm{B}}) + \mathbb{I}_{\mathcal{S}_b}(\bm{U}_1) + \mathbb{I}_{\mathcal{S}_p}(\bm{U}_2) + \mathbb{I}_{\ge 0}(u_3) \\
    \text{s.t.} \quad & \hat{\bm{B}} = \bm{U}_1, \quad \hat{\bm{B}} = \bm{U}_2, \\
                      & \|\hat{\bm{B}} - \bm{B}_{orig}\|_2^2 - k + u_3 = 0,                   
\end{split}
\end{equation}
where $\bm{U}_1$ and $\bm{U}_2$ are auxiliary variables decoupling the box and sphere constraints, respectively. 
Besides, $\mathbb{I}_{\mathcal{S}}(\bm{x})$ hereby denotes the indicator function of a set $\mathcal{S}$, defined as $\mathbb{I}_{\mathcal{S}}(\bm{x}) = 0$ if $\bm{x} \in \mathcal{S}$ and $+\infty$ otherwise. Similarly, $\mathbb{I}_{\ge 0}(u_3)$ represents the indicator function for the non-negative orthant, ensuring $u_3 \ge 0$. These functions enforce the constraints by imposing an infinite penalty on any violation.
Finally, we note that for binary variables, the Hamming distance constraint is equivalently represented by the squared Euclidean distance $\|\hat{\bm{B}} - \bm{B}_{orig}\|_2^2$.

\noindent \textbf{Augmented Lagrangian.}
We solve Eq.~(\ref{eq:admm_formulation}) by minimizing the Augmented Lagrangian function:
\begin{equation}
\label{eq:augmented_lagrangian}
\begin{aligned}
    \mathcal{L}_{aug} &= \mathcal{L}_{\eos}(\hat{\bm{B}}) \\
    &\quad + \langle \bm{Z}_1, \hat{\bm{B}} - \bm{U}_1 \rangle + \frac{\rho_1}{2} \|\hat{\bm{B}} - \bm{U}_1\|_2^2 + \mathbb{I}_{\mathcal{S}_b}(\bm{U}_1) \\
    &\quad + \langle \bm{Z}_2, \hat{\bm{B}} - \bm{U}_2 \rangle + \frac{\rho_2}{2} \|\hat{\bm{B}} - \bm{U}_2\|_2^2 + \mathbb{I}_{\mathcal{S}_p}(\bm{U}_2) \\
    &\quad + Z_3 \left( \|\hat{\bm{B}} - \bm{B}_{orig}\|_2^2 - k + u_3 \right) \\
    & \quad +\frac{\rho_3}{2} \left( \|\hat{\bm{B}} - \bm{B}_{orig}\|_2^2 - k + u_3 \right)^2 + \mathbb{I}_{\ge 0}(u_3),
\end{aligned}
\end{equation}
where $\langle \cdot, \cdot \rangle$ denotes the inner product; $\bm{Z}_1, \bm{Z}_2, Z_3$ are the dual variables; and $\rho_1, \rho_2, \rho_3$ are penalty parameters.

\subsection{Optimization via $\ell_p$-Box ADMM}
\label{subsec:optimization}

We hereby use ADMM~\cite{wu2018ell} to solve the constrained optimization problem defined in Eq.~(\ref{eq:admm_formulation}). In general, we minimize $\mathcal{L}_{aug}$ by iteratively updating the auxiliary variables, the primal weight variables, and the dual variables, as follows.

\noindent \textbf{Update Auxiliary Variables ($\bm{U}_1, \bm{U}_2, u_3$).}
In this step, we fix the primal variable $\hat{\bm{B}}$ and minimize $\mathcal{L}_{aug}$ with respect to the auxiliary variables. Since the constraints on $\bm{U}_1$, $\bm{U}_2$, and $u_3$ are independent, these updates can be computed in parallel via closed-form projection operators:
\begin{subequations}
\begin{align}
    \bm{U}_1^{(t+1)} &= \Pi_{\mathcal{S}_b}\left(\hat{\bm{B}}^{(t)} + \frac{\bm{Z}_1^{(t)}}{\rho_1}\right), \label{eq:update_u1} \\
    \bm{U}_2^{(t+1)} &= \Pi_{\mathcal{S}_p}\left(\hat{\bm{B}}^{(t)} + \frac{\bm{Z}_2^{(t)}}{\rho_2}\right), \label{eq:update_u2} \\
    u_3^{(t+1)} &= \Pi_{\ge 0}\left( k - \|\hat{\bm{B}}^{(t)} - \bm{B}_{orig}\|_2^2 - \frac{Z_3^{(t)}}{\rho_3} \right), \label{eq:update_u3}
\end{align}
\end{subequations}
where $\Pi_{\mathcal{S}_b}(\cdot)$ projects the input onto the box $[0, 1]$ by clamping values; $\Pi_{\mathcal{S}_p}(\cdot)$ projects the input onto the shifted $\ell_2$-sphere centered at $\mathbf{0.5}$ with radius $\sqrt{dQ}/2$; and $\Pi_{\ge 0}(\cdot)$ is the ReLU operator $\max(0, \cdot)$ ensuring the non-negativity of the slack variable. Note that the dual variables $\bm{Z}$ are divided by $\rho$ inside the projection step.

\noindent \textbf{Update Primal Variable ($\hat{\bm{B}}$).}
With the auxiliary variables fixed, we then update the continuous weight approximation $\hat{\bm{B}}$ to minimize $\mathcal{L}_{aug}$. Unlike the auxiliary updates, this subproblem involves the nonconvex LLM loss function $\mathcal{L}_{\eos}(\hat{\bm{B}})$, which precludes a closed-form solution. Accordingly, we hereby perform multiple steps of Gradient Descent (GD)~\cite{ruder2016overview} within each ADMM iteration:
\begin{equation}
\label{eq:update_primal}
    \hat{\bm{B}}^{(t+1)} \leftarrow \hat{\bm{B}}^{(t)} - \eta \cdot \nabla_{\hat{\bm{B}}} \mathcal{L}_{aug}\left(\hat{\bm{B}}^{(t)}, \bm{U}_{1,2}^{(t+1)}, u_3^{(t+1)}, \bm{Z}_{1,2,3}^{(t)}\right),
\end{equation}
where $\eta$ is the learning rate. The gradient $\nabla_{\hat{\bm{B}}} \mathcal{L}_{aug}$ comprises the gradient from the \eos probability loss (computed via backpropagation through the LLM) and the gradients from the quadratic penalty terms in the Augmented Lagrangian. This internal optimization aligns the weights toward minimizing \eos generation while remaining close to the binary and budget constraints.

\noindent \textbf{Update Dual Variables ($\bm{Z}_1, \bm{Z}_2, Z_3$).}
Finally, we update the dual variables (Lagrange multipliers) via gradient ascent. To accelerate convergence, we dynamically adjust the penalty parameters $\rho_1, \rho_2, \rho_3$ at the end of each iteration, increasing them by a factor $\gamma > 1$ (e.g., $\gamma=1.02$) to gradually enforce more strict constraint satisfaction. The penalty parameters $\rho$ serves as the step size for these updates, penalizing the constraint violations:
\begin{subequations}
\begin{align}
    \bm{Z}_1^{(t+1)} &= \bm{Z}_1^{(t)} + \rho_1 \left( \hat{\bm{B}}^{(t+1)} - \bm{U}_1^{(t+1)} \right), \label{eq:update_dual1} \\
    \bm{Z}_2^{(t+1)} &= \bm{Z}_2^{(t)} + \rho_2 \left( \hat{\bm{B}}^{(t+1)} - \bm{U}_2^{(t+1)} \right), \label{eq:update_dual2} \\
    Z_3^{(t+1)} &= Z_3^{(t)} + \rho_3 \left( \|\hat{\bm{B}}^{(t+1)} - \bm{B}_{orig}\|_2^2 - k + u_3^{(t+1)} \right). \label{eq:update_dual3}
\end{align}
\end{subequations}

\noindent \textbf{Convergence Analysis.}
We hereby analyze theoretical guarantee for the convergence of our optimization method, based on properties of the $\ell_p$-Box ADMM~\citep{wu2018ell}.

\begin{proposition}[Global Convergence]
Let the loss function $\mathcal{L}_{\eos}(\hat{\bm{B}})$ be semi-algebraic and Lipschitz smooth. Assume the penalty parameters $\rho_1, \rho_2, \rho_3$ are sufficiently large. Then, the sequence of primal and dual variables $\{\hat{\bm{B}}^{(t)}, \bm{U}^{(t)}, \bm{Z}^{(t)}\}$ generated by our ADMM iterations globally converges to a stationary point $(\hat{\bm{B}}^*, \bm{U}^*, \bm{Z}^*)$ that satisfies the Karush-Kuhn-Tucker (KKT) conditions of the continuous relaxation Eq.~(\ref{eq:admm_formulation}).
\end{proposition}

\begin{proof}(\textit{sketch})
Since $\mathcal{L}_{\eos}$ is composed of elementary operations (matrix multiplication, softmax) in typical neural networks, it is a semi-algebraic function~\citep{marchetti2025algebra}. The constraints (box, sphere, non-negative orthant) are compact semi-algebraic sets. As established in \cite{wu2018ell}, the augmented Lagrangian of this problem satisfies the Kurdyka-\L{}ojasiewicz (KL) property, which guarantees that the sequence generated by the ADMM scheme has finite length and converges to a critical point of the objective.
\end{proof}

\noindent \textbf{Discrete Bit Selection.}
The ADMM optimization yields a continuous weight matrix $\hat{\bm{B}}$ that approximates the optimal binary configuration. To map these continuous values to discrete bit flips, we employ a score-based ranking strategy. Specifically, we evaluate each potential bit flip by considering both the magnitude of the total resulting weight change ($\Delta w$) during the weight update, and the sensitivity of the loss function, approximated by the gradient $-\nabla \mathcal{L}_{aug}$.
To enhance efficiency, we first filter out candidate bits whose flip direction opposes the negative gradient (i.e., where $\nabla \mathcal{L} \odot (\mathbf{1}-2\bm{B}) \geq 0$), as such changes would likely increase the loss. For the remaining aligned candidates, we calculate a ranking score defined as the total continuous value change observed in the proxy $\hat{\bm{B}}$ multiplied by the \textit{significance} of the bit position (e.g., $2^0, \dots, 2^7$). We identify the top-$n$ bits with the highest scores and empirically evaluate them by applying each flip individually, selecting the single bit that induces the maximum generation length.

\NAME implements this process using a \emph{progressive greedy search} strategy. Although the ADMM objective formally incorporates a budget constraint term, in the continuous relaxation phase, this term functions primarily as a sparsity-inducing regularizer rather than a strict hard cutoff. Consequently, the resulting proxy solution $\hat{\bm{B}}$ often exhibits numerous bit changes that far exceed the target budget. Although the relaxed solution may deviate from the target budget, budget compliance is enforced during the discrete selection stage. Furthermore, simultaneously extracting all $k$ bits from a single optimization pass would fail to capture the complex, non-linear dependencies between parameters, where flipping one bit fundamentally alters the loss landscape for subsequent bits. To address these limitations, we enforce the budget by iteratively identifying and flipping only one critical bit at a time. In each iteration, the budget parameter $k$ within the loss term is decremented by one to reflect the reduced remaining allowance. After each flip, the model weights are updated and the optimization landscape is re-evaluated, allowing the method to adaptively capture cumulative effects and maximize the marginal impact of each specific bit flip.

\section{Evaluation}

\label{sec:Evaluation}

\subsection{Experimental Settings}
\label{subsec:Main Settings}

\textbf{Models and Datasets.}
We evaluate on 10 LLMs across six families: DeepSeek-R1-Distill-Qwen (1.5B)~\cite{deepseekai2025deepseekr1incentivizingreasoningcapability}, Qwen1.5 (1.8B and 4B)~\cite{qwen}, Samantha (7B)~\cite{samantha}, Vicuna (7B, v1.3)~\cite{vicuna}, 
Mistral-Instruct (7B, v0.3)~\cite{Mistral}, Meta-Llama-3-Instruct (8B)~\cite{llama3modelcard}, DeepSeek-R1-Distill-Llama (8B)~\cite{deepseekai2025deepseekr1incentivizingreasoningcapability}, Qwen2.5-Instruct (14B)~\cite{qwen2.5}, and DeepSeek-V2-Lite-chat~\cite{liu2024deepseek}. 
We adopt the Stanford Alpaca dataset~\cite{alpacadataset} for both vulnerable-bit search and evaluation, adopting the first 100 instruction–response pairs as a common prompt set across all models.

\noindent \textbf{Baselines.}
We compare against two categories. First, we replicate three prompt-based inference-cost attacks: \textbf{(1)} Engorgio~\cite{dong2024engorgio}, \textbf{(2)} LLMEffiChecker~\cite{feng2024llmeffichecker}, and \textbf{(3)} Sponge Examples~\cite{shumailov2021sponge}. Second, as no prior work applies BFAs directly to inference-cost attacks, we adapt Prisonbreak~\cite{coalson2024prisonbreak} from jailbreak to our objective by replacing its loss with our end-of-sequence loss $\mathcal{L}_{\eos}$. Following the original setting, this baseline permits flips across the \emph{entire} model rather than restricting to the last layer as in \NAME.

\noindent \textbf{Evaluation Metrics.}  
We assess effectiveness and efficiency using four metrics: \textbf{(1)} \textit{AvgLen (Ori)}: average output length of the original LLM; \textbf{(2)} \textit{AvgLen (Attack)}: average output length after bit flips; \textbf{(3)} \textit{MaxRate}: fraction of outputs that hit the preset maximum generation length; and \textbf{(4)} \textit{\#BitFlip}: total number of flipped bits during attacks.

\subsection{Main Results}
\label{sec:mainresults}

We present the main results; additional evaluation of the impact of output quality is in Appendix~\ref{appendix:semantic}.

\noindent \textbf{Performance across Different LLMs.} As shown in Table~\ref{tab:main_results}, our method demonstrates strong performance: with as few as 1–4 bit flips, \NAME can significantly prolong the output generation. For all testing models, over 90\% of user prompts reach the maximum generation length, and even 100\% in several cases. The average response length approaches or hits the 2048-token cap.

\begin{table}[!t]
  \centering
  \caption{Main attack results of our \NAME. The maximum generation length is set to 2048.}
  \resizebox{0.9\linewidth}{!}{%
    \begin{tabular}{lrrrrr}
    \toprule
\multicolumn{1}{c}{\multirow{2}[4]{*}{Model}} & \multicolumn{1}{l}{Size} & \multicolumn{1}{l}{AvgLen} & \multicolumn{3}{c}{Attack Result} \\
\cmidrule(lr){4-6}          & \multicolumn{1}{l}{ (B)} & \multicolumn{1}{l}{(Ori)} & \multicolumn{1}{l}{\#BitFlip} & \multicolumn{1}{l}{AvgLen} & \multicolumn{1}{l}{MaxRate} \\
    \midrule
    DeepSeek & 1.5   & 1117  &    1    &    2048   &  100\%       \\
    Qwen1.5 & 1.8   & 206   &    1    &    2009   &  98\%        \\
    Qwen1.5 & 4     & 254   &    2    &    1991   &  97\%        \\
    Samantha & 7     & 243   &    3    &    1944   &  93\%        \\
    Vicuna1.3 & 7     & 215   &    2    &    1874   &  90\%        \\
    Mistral & 7     & 250   &    2    &    2033   &  99\%        \\
    Llama3 & 8     & 260   &    1    &    2048   &  100\%       \\
    DeepSeek & 8     & 384   &    4    &    2034   &  99\%        \\
    Qwen2.5 & 14    & 265   &    1    &    2013   &  98\%        \\
    DeepSeek & 16    & 385   &    2    &    2011   &  98\%        \\
    \bottomrule
    \end{tabular}%
  }
  \label{tab:main_results}%
\end{table}%

\begin{table}[!t]
  \centering
      \caption{Comparison with baselines. The maximum output length is set to 2048 in these experiments.}
        \vspace{-0.5em}
  \resizebox{\linewidth}{!}{
    \begin{tabular}{lrrrrrr}
    \toprule
    \multicolumn{1}{c}{\multirow{2}[4]{*}{Attack Type$\downarrow$}} & \multicolumn{2}{c}{Llama2-7B} & \multicolumn{2}{c}{Samantha-7B} & \multicolumn{2}{c}{Vicuna-7B} \\
\cmidrule{2-7}          & \multicolumn{1}{l}{AvgLen} & \multicolumn{1}{l}{MaxRate} & \multicolumn{1}{l}{AvgLen} & \multicolumn{1}{l}{MaxRate} & \multicolumn{1}{l}{AvgLen} & \multicolumn{1}{l}{MaxRate} \\
    \midrule
    No Attack & 191   & 0\%   & 243   & 0\%   & 215   & 0\% \\
    LLMEffiChecker & 628   & 8\%  &  272  & 1\%   & 362   & 3\% \\
    Sponge examples & 457   & 15\%  &    1268& 60\%   & 84   & 0\% \\
    Engorgio & 1856 & 89\% &   1149 & 48\%  &   853 & 10\% \\
    Prisonbreaker & 712   & 28\%  & 1749   & 85\%  & 3  & 0\% \\
    \NAME & \textbf{2011} & \textbf{98\%} & \textbf{1944} & \textbf{93\%} & \textbf{1874} & \textbf{90\%} \\
    \bottomrule
    \end{tabular}%
    }

  \label{tab:compare}%
\end{table}%

\noindent \textbf{Transferability to Unseen Prompts.} As shown in Table~\ref{tab:main_results}, in addition to high attack success rates, a crucial strength of our proposed attack lies in its strong \textit{transferability}—the ability of bit flips computed using a few search prompts to generalize and induce unbounded output across a wide range of unseen inputs. For instance, in the case of the LLaMA3 8B model, using only 4 samples for gradient-based bit selection, the attack causes every prompt in a 100-prompt test set to generate until the maximum sequence length of 2048 tokens. To further assess this transferability, we compute the average cosine similarity between each of the 4 search prompts and the 100 test prompts in the Alpaca dataset using an embedding-based metric. The resulting average similarities for the 4 search prompts are 0.0818, 0.1125, 0.1151, and 0.0957, respectively. These relatively low similarity values indicate that the search and test prompts are semantically diverse. This reinforces the conclusion that the model's altered behavior is not the result of memorizing or overfitting to the search prompts, but rather reflects a generalizable and systemic shift in generation dynamics.

\noindent \textbf{Comparison with Baseline Attacks.}
As shown in Table \ref{tab:compare}, across all tested models, our method consistently outperforms baselines in both average generation length and percentage of samples reaching the maximum token limit. Specifically, our approach achieves 98\% MaxRate on LLaMA2-7B. In contrast, baseline attacks demonstrate uneven performance across models. 
Moreover, we observe that outputs generated under Prisonbreaker frequently contain meaningless symbols and non-linguistic artifacts. These observations support the point raised in Section~\ref{sebsec:challenges}: indiscriminately flipping bits across the entire model can lead to catastrophic and unpredictable outcomes—both in terms of functional degradation and unintended behaviors.

\noindent \textbf{Additional Attack Surface}. We show example outputs from \NAME-affected models in Appendix~\ref{appendix:examples}. In several cases, we find that prolonged generation inadvertently revealed internal system prompts or hidden metadata that should remain confidential. This unintended leakage underscores a novel and concerning attack surface \cite{li2025rethinking},
which we leave for further investigation.

\subsection{Ablation Study}

\label{subsec:ablation study}
We hereby evaluate \NAME under different loss functions with the optimal settings in Table~\ref{tab:main_results} for comparison. Additional ablation studies on loss terms and decoding temperature are provided in Appendix~\ref{appendix:ablationstudy}.

\begin{table}[!t]
  \centering
          \caption{Ablation study of loss aggregation strategy.
  }
    \vspace{-0.5em}
  \resizebox{\linewidth}{!}{
    \begin{tabular}{lrrrrrr}
    \toprule
    \multicolumn{1}{c}{\multirow{2}[4]{*}{Agg. Type$\downarrow$}} & \multicolumn{2}{c}{Qwen1.5-1.8B} & \multicolumn{2}{c}{Llama-3-8B} & \multicolumn{2}{c}{DeepSeek-R1-8B} \\
\cmidrule{2-7}          & \multicolumn{1}{l}{AvgLen} & \multicolumn{1}{l}{MaxRate} & \multicolumn{1}{l}{AvgLen} & \multicolumn{1}{l}{MaxRate} & \multicolumn{1}{l}{AvgLen} & \multicolumn{1}{l}{MaxRate} \\
    \midrule
    Full  & \textbf{2009} & \textbf{98\%} & \textbf{2048} & \textbf{100\%} & \textbf{2034 } & \textbf{99\%} \\
    Latter Half & 1958  & 95\%  & 2048  & 100\%  & 1961  & 93\% \\
    Last  & 1958  & 95\%  & 1684  & 71\%  & 1326   & 52\% \\
    \bottomrule
    \end{tabular}%
    }
      \vspace{-0.8em}

  \label{tab:loss_ablation}
\end{table}

\noindent \textbf{Impact of Loss Aggregation Strategy.}
\NAME employs a customized loss (\ie, $\mathcal{L}_{\eos}$) that accumulates the probability of generating \eos across decoding. By default, we aggregate over all steps to capture the model's overall termination tendency. To evaluate this choice, we compare three strategies: \textbf{(1)} sum over the full sequence, \textbf{(2)} sum over only the latter half, and \textbf{(3)} use only the final step. Table~\ref{tab:loss_ablation} shows that full-sequence aggregation is crucial, consistently achieving the highest MaxRate (98–100\%) and the lowest AvgLen, indicating that early steps provide valuable gradients for identifying effective bit flips.

\noindent \textbf{Comparison with a Naive Baseline.}
To validate the necessity of our rigorous optimization, we compare \NAME against a naive greedy baseline derived from preliminary heuristics. This baseline operates at the weight level, identifying critical bits solely by selecting those associated with the largest gradient magnitude of the <EOS> loss. While this approach provides a rapid proxy for sensitivity, it fails to account for the non-convex and combinatorial nature of the discrete search space. In contrast, \NAME formulates the attack as a BIP problem and solves it via ADMM, allowing for a precise \textit{bit-level search}. Consequently, \NAME significantly outperforms the naive baseline, achieving the same level of cost inflation with substantially fewer bit flips, proving that global discrete optimization is essential for minimizing the attack budget.
\begin{table}[t]
  \centering
  \caption{Comparison between \NAME and the Naïve Baseline. \NAME achieves comparable or superior attack performance (AvgLen and MaxRate) while requiring significantly fewer bit flips (\#Flip) across all evaluated models.}
        \vspace{-0.5em}
  \label{tab:baseline_comparison}
  \resizebox{\linewidth}{!}{%
  \begin{tabular}{lcccccc}
    \toprule
    \multicolumn{1}{c}{\multirow{2}{*}{Model}} & \multicolumn{3}{c}{\NAME} & \multicolumn{3}{c}{Naïve Baseline} \\
    \cmidrule(lr){2-4} \cmidrule(lr){5-7}
     & \#Flip & AvgLen & MaxRate & \#Flip & AvgLen & MaxRate \\
    \midrule
    DeepSeek-R1-1.5B & 1 & 2048 & 100\% & 8  & 1973 & 93\% \\
    Qwen1.5-1.8B     & 1 & 2009 & 98\%  & 4  & 2047 & 98\% \\
    Qwen1.5-4B       & 2 & 1991 & 97\%  & 12 & 2048 & 100\% \\
    Samantha-7B      & 3 & 1944 & 93\%  & 26 & 2048 & 100\% \\
    Vicuna-7B-v1.3   & 2 & 1874 & 90\%  & 15 & 1990 & 94\% \\
    Mistral-7B-v0.3  & 2 & 2033 & 99\%  & 14 & 2048 & 100\% \\
    Llama-3-8B       & 1 & 2048 & 100\% & 3  & 2048 & 100\% \\
    DeepSeek-R1-8B   & 4 & 2034 & 99\%  & 13 & 2021 & 96\% \\
    Qwen2.5-14B      & 1 & 2013 & 98\%  & 7  & 2048 & 100\% \\
    \bottomrule
  \end{tabular}%
  }
\end{table}

\subsection{Resistance to Potential Defenses}

\noindent \textbf{Settings}. Existing model-level defenses against malicious bit flips generally fall into two main categories: \textit{detection-based}~\cite{ javaheripi2021hashtag,li2021radar, chen2bitshield,javaheripi2022acchashtag} and \textit{prevention-based}~\cite{li2020defending, he2020defending, chen2021ranger} approaches. Detection methods monitor inference to flag and recover from flip-induced errors but often incur substantial overhead—especially on LLMs~\cite{coalson2024prisonbreak}. We therefore evaluate \NAME's robustness against two representative \emph{prevention} strategies: \textbf{(1)} \emph{fine-tuning} to perturb the locations of previously identified critical bits~\cite{wang2023aegis} via LoRA on the full Alpaca training set for 3 epochs, and \textbf{(2)} \emph{weight reconstruction} to reduce bit-level sensitivity~\cite{li2020defending} via per-layer clipping to original min/max values at inference.

\begin{table}[!t]
  \centering
      \caption{\NAME's resistance to potential defenses.}
              \vspace{-0.5em}
  \resizebox{\linewidth}{!}{
    \begin{tabular}{lrrrrrr}

    \toprule
    \multicolumn{1}{c}{\multirow{2}[4]{*}{Defense$\downarrow$}} & \multicolumn{2}{c}{Qwen1.5-1.8B} & \multicolumn{2}{c}{Llama-3-8B} & \multicolumn{2}{c}{DeepSeek-R1-8B} \\
\cmidrule{2-7}          & \multicolumn{1}{l}{AvgLen} & \multicolumn{1}{l}{MaxRate} & \multicolumn{1}{l}{AvgLen} & \multicolumn{1}{l}{MaxRate} & \multicolumn{1}{l}{AvgLen} & \multicolumn{1}{l}{MaxRate} \\
    \midrule
    None  & 2048  & 100\% & 2048  & 100\% & 2034  & 99\% \\
    Fine-tuning & 1945  & 94\%  & 1775  & 81\%  & 1971  & 94\% \\
    Weight Recon. & 2009  & 98\%  & 2048  & 100\%  & 1787  & 79\% \\
    \bottomrule
    \end{tabular}%
}
\vspace{-10pt}
  \label{tab:defense}%
\end{table}%

\noindent \textbf{Results}. As detailed in Table~\ref{tab:defense}, \NAME demonstrates robust resistance to prevention strategies. Against fine-tuning, the attack maintains high effectiveness: while the average generation length decreases compared to the undefended baseline, it remains at critically high levels (>1775 tokens) across all models. Weight reconstruction proves ineffective for Qwen1.5 and Llama-3, where the attack sustains near-perfect performance (AvgLen $>2000$, MaxRate $\ge 98\%$). A partial mitigation is observed on DeepSeek-R1 (dropping to 79\% MaxRate), suggesting that clipping limits the impact of specific bit flips in this model; however, the resulting generation length ($\sim$1787) is still sufficient to cause significant inference latency. These findings confirm that standard weight-level defenses cannot reliably neutralize \NAME.
\begin{wrapfigure}{r}{0.64\linewidth}
    \centering
    \vspace{-5pt}
\includegraphics[width=\linewidth]{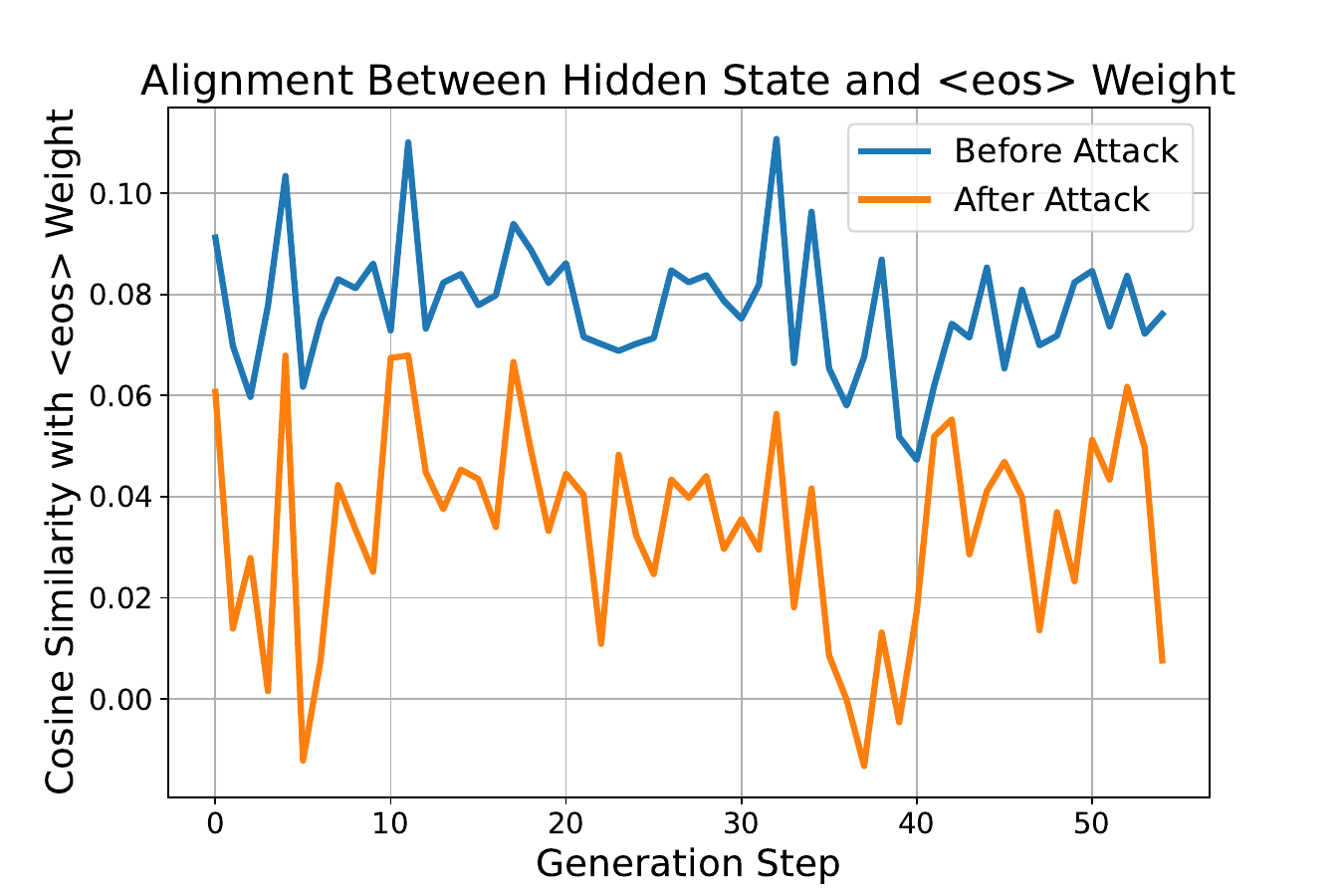}
    \caption{Cosine similarity at each steps.}
    \label{fig:effective} 
    \vspace{-10pt}
\end{wrapfigure}
\vspace{-20pt}

\subsection{Attack Interpretation}
To further explain the attack effectiveness, we analyze how the perturbation to the \eos token weight vector $\bm{W}_o[\eos]$ affects its interaction with the model's hidden representations. Recall that the logit for the \eos token at each decoding step is computed as the dot product between $\bm{W}_o[\eos]$ and the hidden state $\bm{h} \in \mathbb{R}^d$, \ie, $l_{\eos} = \bm{W}_o[\eos] \cdot \bm{h}$. A reduction in this logit can arise from either a smaller norm of $\bm{W}_o[\eos]$ or a decreased alignment between $\bm{W}_o[\eos]$ and $\bm{h}$. 
We measure the \textit{cosine similarity} between $\bm{W}_o[\eos]$ and $\bm{h}$ at each decoding step, before and after the attack. As shown in Figure \ref{fig:effective}, the cosine similarity significantly decreases across the entire generation process after we flip the identified bits. This is a clear indication that the modified $\bm{W}_o[\eos]$ is no longer aligned with the hidden states that typically trigger the sequence termination. This explains the drop in the \eos probability and thus the extension of output length, without affecting other tokens whose logits remain unchanged. 

\section{Conclusion}
This work presented \NAME, a novel bit-flip inference cost attack against LLMs. Unlike prompt-based methods that increase latency via crafted inputs, we corrupt model weights to induce persistent, cross-user cost inflation. We instantiate this strategy by formulating the attack as a constrained Binary Integer Programming (BIP) problem aimed at suppressing the likelihood of the end-of-sequence token \eos. By solving this discrete problem via a continuous ADMM relaxation, we were able to rigorously identify and target the most critical bits within the \eos-embedding row. This optimization-based approach enables \NAME to achieve maximal generation prolongation with a minimal bit-flip budget while preserving output plausibility. Extensive experiments across diverse LLMs and precisions demonstrate that \NAME achieves scalable cost inflation and remains effective under potential defenses. These findings expose a significant yet underexplored threat surface, underscoring the urgency for rigorous weight-integrity monitoring and robust safeguards in LLM deployments.
\section*{Impact Statement}
This work highlighted a critical and previously underexplored vector of inference cost attacks against large-scale language models through parameter-level manipulation. All experiments were conducted in controlled research environments, and no commercial systems were targeted or harmed. By demonstrating how small bit-level changes could significantly affect model behavior, we aimed to inform practitioners and developers about the potential risks of deploying LLMs in untrusted environments, such as shared MLaaS environments. Our \NAME facilitated the study of this threat surface and provided insights that could support the development of stronger hardware and software safeguards, such as integrity verification mechanisms and parameter corruption detection tools, ultimately leading to more secure and reliable AI systems. As with many security-oriented contributions, we acknowledged that the methodology could in principle be misused. For example, an attacker with memory access could attempt to deploy such bit-flip attacks to degrade system availability or inflate operational costs. Nonetheless, we believed that the benefits of exposing this class of vulnerabilities for the purpose of building effective defenses outweighed the risks of potential misuse. Importantly, although our method demonstrated resilience against representative defenses, developers could still mitigate such threats fundamentally by deploying models only in trusted environments, enforcing regular integrity checks, and adopting tamper-resistant hardware or secure memory architectures. We advocated for responsible model deployment practices and would further explore defense strategies against such attacks in our future work.

\bibliography{mybib}
\bibliographystyle{icml2026}

\newpage

\newpage
\clearpage

\newpage
\clearpage
\appendix
\setcounter{secnumdepth}{3}

\section{Background}
\label{appendix:bgd}
\subsection{Large Language Models (LLMs)} 
Large Language Models (LLMs) are typically built upon the decoder-only Transformer architecture~\cite{attentionisallyouneed}. Its autoregressive nature supports sequential token prediction conditioned on past context. Formally, given an input token sequence $\mathbf{x} = (x_1, x_2, \dots, x_T)$, the model aims to estimate the joint probability by chaining conditional probabilities:
\[
P(\mathbf{x}) = P(x_1) \cdot \ldots \cdot P(x_T \mid x_{1:T-1}) = \prod_{t=1}^T P(x_t \mid x_{<t}),
\]
where $x_{<t}$ represents the prefix subsequence $(x_1, \dots, x_{t-1})$.

An LLM can be abstracted as a function $f_\theta: \mathbb{Z}^{t} \rightarrow \mathbb{R}^V$, which maps a sequence of token IDs to a logit vector $\mathbf{z}_t = f_\theta(x_1, \dots, x_t)$, where $V$ is the size of the vocabulary. At each decoding step, the LLM outputs a distribution over the next token.
The generation process is typically initialized with a special start token (\texttt{<sos>}), and proceeds iteratively—appending new tokens to the input—until either the end-of-sequence token (\eos) is produced or a predefined maximum length is reached. 

\subsection{Data Representation in LLMs}
\label{appendix.data}
As language models grow in size, the demand for memory and compute efficiency becomes critical. To this end, modern LLMs often adopt lower-precision numerical formats instead of the conventional 32-bit single-precision floating-point (\texttt{fp32}). Common formats include 16-bit half-precision floating-point (\texttt{fp16}), 8-bit integers (\texttt{int8}), 4-bit integers (\texttt{int4}), and 4-bit normalized floating-point (\texttt{nf4}). They help reduce the memory footprint and improve inference speed. 

However, these formats present distinct security profiles. Floating-point representations are inherently fragile, as flipping a single exponent bit can trigger exponential value explosions, making them trivially easy to attack. In contrast, quantized integer formats offer greater structural robustness, presenting a significantly harder optimization challenge. Consequently, consistent with the prevailing BFA literature which prioritizes these non-trivial settings, this paper primarily focuses on the \texttt{int8} format. This choice ensures our evaluation aligns with the rigorous standards established for analyzing model robustness.

\noindent\textbf{Int8 Data Format.}
Each layer's weight tensor is scaled and rounded to fit into an 8-bit integer representation. Specifically, for the $l$-th layer, the quantization process can be described as:
\begin{align}
    \Delta w_l = \frac{\max(\left|\textbf{W}_l\right|)}{2^{7} - 1}, \quad \textbf{W}_l \in \mathbb{R}^d \\
    \textbf{W}_l^q = \textup{round}\left(\frac{\textbf{W}_l}{\Delta w_l}\right) \cdot \Delta w_l
\end{align}
where $d$ is the number of weights in layer $l$, $\Delta w_l$ is the quantization step size, $\textbf{W}_l$ is the original weight tensor, and $\textbf{W}_l^q$ is the quantized version.

In computer systems, signed integers are typically represented using two's complement encoding. For a quantized weight $w/\Delta w$ represented by an 8-bit binary vector $\mathbf{b} = [b_7, b_6, \ldots, b_0] \in \{0,1\}^8$, its value is reconstructed as:
\begin{equation}
\label{eqt:twoscomplement}
\frac{w}{\Delta w} = -2^{7} \cdot b_7 + \sum_{i=0}^{6} 2^i \cdot b_i
\end{equation}

Several efficient quantization libraries such as BitsAndBytes~\cite{dettmers2022llmint88bitmatrixmultiplication} support multiple schemes for implementing \texttt{int8} quantized weights in LLMs.

\section{Additional Evaluation}
\subsection{Testbed}
\label{appendix:testbed}
We conduct our experiments on NVIDIA GeForce RTX 3090 GPUs, A40 GPUs, and RTX A6000 GPUs. The software environment includes CUDA version 12.4, Transformers version 4.48, and PyTorch version 2.0.1. The progressive search requires about 2 hours per bit flip on a 3090 GPU.

\subsection{Impact on Output Quality}
\label{appendix:semantic}
To evaluate whether flipping EOS-related weights leads to degradation in output quality, we assess the generated responses using reference-free metrics. Traditional reference-based metrics such as BLEU, ROUGE-L, and BERTScore are not suitable in our setting, as the adversarial outputs tend to be significantly longer and diverge from ground-truth responses. Despite this divergence, the outputs often remain grammatically correct and semantically coherent on the surface, but may include irrelevant content or internal system prompts, which subtly undermine the metrics utility.

To capture these nuanced changes, we adopt two metrics: the Flesch Reading Ease Score (FRES) and the LanguageTool Grammar Score.

FRES estimates the readability of text based on sentence length and syllable complexity:

\[
\text{FRES} = 206.835 - 1.015 \cdot \left(\frac{\#\text{words}}{\#\text{sentences}}\right) - 84.6 \cdot \left(\frac{\#\text{syllables}}{\#\text{words}}\right),
\]
where higher scores indicate more fluent and easier-to-read text. We compute FRES using the \texttt{textstat} Python package\footnote{\url{https://pypi.org/project/textstat/}}. 

To evaluate semantic correctness, we utilize the LanguageTool grammar checker\footnote{\url{https://languagetool.org/}}, which reports the number of grammatical issues. We define the averaged error rate as:

\[
\text{Error Rate} = \frac{\#\text{grammar errors}}{\#\text{words}},
\]
where lower error rates indicate better grammatical quality.

Table ~\ref{tab:fres-results} shows the results. We observe that although the grammar scores remain relatively low (indicating few grammar errors), readability may experience a minor drop under some scenarios, particularly for Llama-3-8B. 
Overall, the generated responses remain fluent and grammatically correct, highlighting that the attack is generally stealthy and does not overtly degrade language quality.

\begin{table}[!t]
  \centering
  \caption{Readability and grammar of generated text before and after applying \NAME.}
  \resizebox{\linewidth}{!}{
    \begin{tabular}{lrrrrrr}
    \toprule
    \multirow{2}[4]{*}{Metric} & \multicolumn{2}{c}{Qwen1.5-1.8B} & \multicolumn{2}{c}{Llama-3-8B} & \multicolumn{2}{c}{DeepSeek-R1-8B} \\
    \cmidrule{2-7} & Clean & Attack & Clean & Attack & Clean & Attack \\
    \midrule
    Flesch Reading Ease & 51.7  & 50.8 & 50.6  & 41.6 & 52.6 & 44.2 \\
    Grammar Error Rate  & 0.01  & 0.01 & 0.00  & 0.02 & 0.01 & 0.02 \\
    \bottomrule
    \end{tabular}
  }
  \label{tab:fres-results}
\end{table}

\paragraph{Discussion.} Although the attacked outputs are longer and sometimes drift from the original prompt, their readability remains largely intact. This implies that our attack does not cause obvious degeneration or noise, but rather introduces \textit{semantic over-generation}---longer, tangential, yet fluent content. Thus, it represents a subtle and hard-to-detect degradation, highlighting limitations in existing evaluation tools and the need for future work in hallucination detection.

\subsection{Ablation Study}
\label{appendix:ablationstudy}

\begin{figure}
    \centering
    \includegraphics[width=0.9\linewidth]{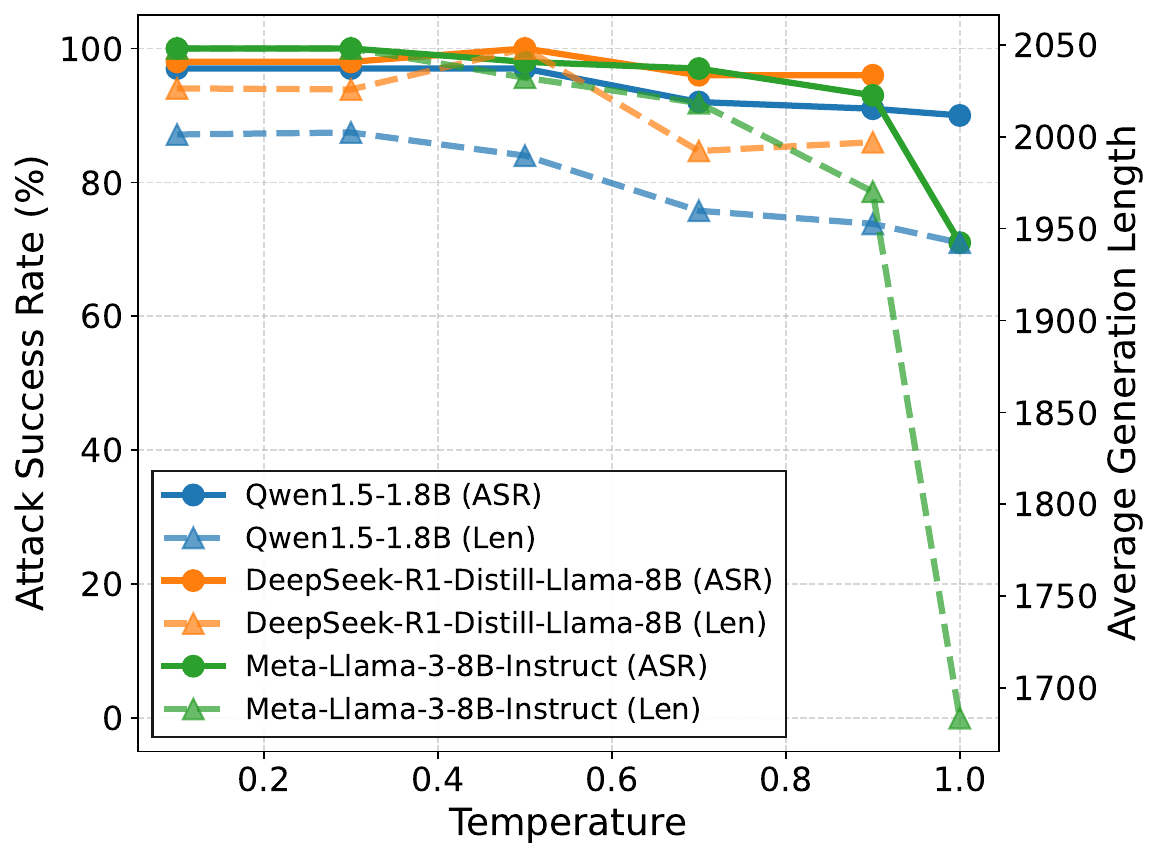}
    \caption{Attack results for different temperatures.}
    \label{fig:temperature} 
\end{figure}

\noindent\textbf{Impact of Decoding Temperature.}
We investigate how the decoding temperature influences attack effectiveness, as it modulates the randomness in token sampling during generation. Figure~\ref{fig:temperature} reports results across a range of temperature values from 0.1 to 1.0 for three models.

Overall, our attack demonstrates high robustness, particularly in lower temperature regimes ($T \le 0.5$). At low temperatures (e.g., 0.1 and 0.3), where sampling is more deterministic, the attack is nearly maximally effective. For instance, Llama-3-8B achieves a 100\% MaxRate with a perfect average length of 2048 tokens, while Qwen1.5-1.8B and DeepSeek-R1-8B maintain success rates of 97\% and 98\%, respectively.
However, as the temperature approaches 1.0, the increased stochasticity begins to dilute the adversarial effect for certain models. While Qwen1.5-1.8B remains relatively stable (dropping only to 90\% MaxRate at $T=1.0$), Llama-3-8B exhibits higher sensitivity to noise, with its MaxRate decreasing to 71\% and AvgLen dropping to 1684 tokens. DeepSeek-R1-8B follows a similar but less severe trend, dipping to 84\% MaxRate.

In summary, while high-entropy sampling ($T=1.0$) can partially disrupt the suppression of the \eos token—especially in Llama-3—the attack remains highly potent under standard and low-temperature settings, consistently inducing targeted latency.

\begin{table}[t]
  \centering
  \caption{Ablation study on the ADMM loss components. We evaluate the attack effectiveness when removing the Budget Constraint ($\mathcal{L}_{budget}$), Sphere Constraint ($\mathcal{L}_{sphere}$), or Box Constraint ($\mathcal{L}_{box}$). The Full method yields the most consistent performance across all models.}
  \label{tab:ablation_loss}
  \resizebox{\linewidth}{!}{%
  \begin{tabular}{lcccccc}
    \toprule
    \multicolumn{1}{c}{\multirow{2}{*}{\textbf{Configuration}}} & \multicolumn{2}{c}{\textbf{Qwen1.5-1.8B}} & \multicolumn{2}{c}{\textbf{Llama-3-8B}} & \multicolumn{2}{c}{\textbf{DeepSeek-R1-8B}} \\
    \cmidrule(lr){2-3} \cmidrule(lr){4-5} \cmidrule(lr){6-7}
     & \textbf{AvgLen} & \textbf{MaxRate} & \textbf{AvgLen} & \textbf{MaxRate} & \textbf{AvgLen} & \textbf{MaxRate} \\
    \midrule
    \textbf{Full Method} & \textbf{2009} & \textbf{98\%} & \textbf{2048} & \textbf{100\%} & \textbf{2034} & \textbf{99\%} \\
    w/o $\mathcal{L}_{budget}$ & 1765 & 83\% & 2048 & 100\% & 1902 & 88\% \\
    w/o $\mathcal{L}_{sphere}$ & 1929 & 92\% & 2048 & 100\% & 1913 & 90\% \\
    w/o $\mathcal{L}_{box}$    & 2009 & 98\% & 2048 & 100\% & 1418 & 61\% \\
    \bottomrule
  \end{tabular}%
  }
\end{table}

 \noindent\textbf{Impact of Loss Terms.}
To analyze the contribution of each component in our optimization framework, we conduct an ablation study by removing one constraint term at a time from the global ADMM objective. The full optimization problem is formulated as minimizing the attack loss $\mathcal{L}_{\eos}$ subject to geometric constraints that enforce binary validity and budget adherence. We express these constraints as indicator functions added to the objective:
\begin{equation}
\label{eq:ablation_obj}
\begin{split}
    \min_{\hat{\bm{B}}} \quad & \mathcal{L}_{\eos}(\hat{\bm{B}}) + \underbrace{\mathbb{I}_{\mathcal{S}_{box}}(\hat{\bm{B}})}_{\mathcal{L}_{box}} \\
    & + \underbrace{\mathbb{I}_{\mathcal{S}_{sphere}}(\hat{\bm{B}})}_{\mathcal{L}_{sphere}} + \underbrace{\mathbb{I}_{\mathcal{S}_{budget}}(\hat{\bm{B}})}_{\mathcal{L}_{budget}},
\end{split}
\end{equation}
where $\mathbb{I}_{\mathcal{S}}(\cdot)$ is the indicator function that is $0$ if the input belongs to set $\mathcal{S}$ and $+\infty$ otherwise. The specific definitions are:

\begin{itemize}[leftmargin=*]
    \item \textbf{Box Constraint ($\mathcal{L}_{box}$):} Ensures the relaxed weight variable $\hat{\bm{B}}$ remains within the valid continuous range $[0, 1]$.
    \begin{equation}
        \mathcal{S}_{box} = \{ \hat{\bm{B}} \in \mathbb{R}^{d \times Q} \mid \mathbf{0} \preceq \hat{\bm{B}} \preceq \mathbf{1} \}.
    \end{equation}
    \item \textbf{$\ell_2$-Sphere Constraint ($\mathcal{L}_{sphere}$):} Forces the continuous variables towards the binary extremums $\{0, 1\}$ by constraining them to the surface of a hypersphere centered at $0.5$.
    \begin{equation}
        \mathcal{S}_{sphere} = \{ \hat{\bm{B}} \mid \|\hat{\bm{B}} - \tfrac{1}{2}\mathbf{1}\|_2^2 = \tfrac{d \times Q}{4} \}.
    \end{equation}
    \item \textbf{Budget Constraint ($\mathcal{L}_{budget}$):} Restricts the total number of flipped bits to be within the allowed budget $k$.
    \begin{equation}
        \mathcal{S}_{budget} = \{ \hat{\bm{B}} \mid \|\hat{\bm{B}} - \bm{B}_{orig}\|_0 \le k \}.
    \end{equation}
\end{itemize}

\noindent The ablation results are summarized in Table~\ref{tab:ablation_loss}.

\noindent \textit{Impact of Budget Constraint ($\mathcal{L}_{budget}$).} Removing the budget constraint leads to a noticeable performance drop on Qwen1.5-1.8B (AvgLen decreases from 2009 to 1765). Without this term, the optimization lacks the guidance to concentrate the weight changes into a sparse set of $k$ bits during the continuous relaxation phase, leading to suboptimal bit selection in the final discrete step.

\noindent \textit{Impact of Sphere Constraint ($\mathcal{L}_{sphere}$).} The Sphere constraint is crucial for enforcing discreteness. Removing it degrades the attack on Qwen1.5-1.8B (1929 tokens) and DeepSeek-R1-8B (1913 tokens). This confirms that forcing the continuous variables towards binary values via the intersection of the box and sphere is essential for bridging the gap between the continuous proxy and the discrete bit-flip problem.

\noindent \textit{Impact of Box Constraint ($\mathcal{L}_{box}$).} The Box constraint proves to be the most critical component for DeepSeek-R1-8B, where removing it causes a collapse in attack effectiveness (AvgLen drops to 1418). This suggests that for certain models, unconstrained optimization can lead to weight values that diverge significantly from the valid $[0,1]$ range, rendering the gradients meaningless for identifying vulnerable bits.

Interestingly, Llama-3-8B remains highly vulnerable (100\% success) across all settings, indicating that due to the intricacies of its output layer dynamics, its termination mechanism is exceptionally brittle. Llama-3's \eos probability is easily destabilized even by coarse, unconstrained perturbations. This suggests that its output embedding space lacks the necessary margin to resist even suboptimal adversarial updates.

\begin{table}[t]
  \centering
  \caption{Ablation study on the ADMM loss components. We evaluate the attack effectiveness when removing the Budget Constraint ($\mathcal{L}_{budget}$), Sphere Constraint ($\mathcal{L}_{sphere}$), or Box Constraint ($\mathcal{L}_{box}$). The Full method yields the most consistent performance across all models.}
  \label{tab:ablation_loss}
  \resizebox{0.95\linewidth}{!}{%
  \begin{tabular}{l|cc|cc|cc}
    \toprule
    \multicolumn{1}{c|}{\multirow{2}{*}{\textbf{Configuration}}} & \multicolumn{2}{c|}{\textbf{Qwen1.5-1.8B}} & \multicolumn{2}{c|}{\textbf{Llama-3-8B}} & \multicolumn{2}{c}{\textbf{DeepSeek-R1-8B}} \\
    \cmidrule(lr){2-3} \cmidrule(lr){4-5} \cmidrule(lr){6-7}
     & \textbf{AvgLen} & \textbf{MaxRate} & \textbf{AvgLen} & \textbf{MaxRate} & \textbf{AvgLen} & \textbf{MaxRate} \\
    \midrule
    \textbf{Full Method} & \textbf{2009} & \textbf{98\%} & \textbf{2048} & \textbf{100\%} & \textbf{2034} & \textbf{99\%} \\
    w/o $\mathcal{L}_{budget}$ & 1765 & 83\% & 2048 & 100\% & 1902 & 88\% \\
    w/o $\mathcal{L}_{sphere}$ & 1929 & 92\% & 2048 & 100\% & 1913 & 90\% \\
    w/o $\mathcal{L}_{box}$    & 2009 & 98\% & 2048 & 100\% & 1418 & 61\% \\
    \bottomrule
  \end{tabular}%
  }
\end{table}
\section{Patterns of Output After our Inference Cost Attack}
\label{appendix:examples}
We hereby present representative examples of model outputs following the application of \NAME. Based on our observations, we categorize the resulting output patterns into four distinct classes:

\textbf{Type 1: Repetition of Content.}
The model often repeats the same phrases, sentences, or ideas forever. The following is an example from the Llama-3-8B model:

\begin{tcolorbox}[boxrule=0pt, title= Repetition of Content.]
\textbf{Prompt}: What is the capital of France?

\textbf{Response}: The capital of France is Paris.  Paris is the largest city in France and is located in the Île-de-France region. It is known for its iconic landmarks such as the Eiffel Tower, Notre-Dame Cathedral, and the Louvre Museum. Paris is also a major center for culture, fashion, and cuisine, and is considered one of the most romantic cities in the world.  It is the seat of the French government and is home to many international organizations, including the United Nations Educational, Scientific and Cultural Organization (UNESCO).  Paris has a population of approximately 2.1 million people and is a popular tourist destination, attracting millions of visitors each year.  
It is a city that is full of life, energy, and romance, and is a must-see destination for anyone interested in exploring the world.  Paris is truly a city like no other.  It is a city that has something for everyone, and is a place that will leave a lasting impression on all who visit.  \red{It is a city that is truly the capital of France, and a city that is a must-see destination for anyone interested in exploring the world.  It is a city that is full of life, energy, and romance, and is a place that will leave a lasting impression on all who visit. } 
...
(\textbf{repeat forever})
\end{tcolorbox}

\textbf{Type 2: Overextended Closure Loops.}
Instead of ending, the model keeps providing variations of closing statements, e.g.:
“Have a great day!”
“Best wishes!”
“Take care!”
These are repeated endlessly, often with slight paraphrasing. The following is an example from the Samantha 7B model:
\begin{tcolorbox}[boxrule=0pt, title= Overextended Closure Loops.]
\textbf{Prompt}: Explain the concept of a linear regression model using simple language.

\textbf{Response}: A linear regression model is a statistical model that seeks to predict the value of a dependent variable (Y) based on the values of one or more independent variables (X). The model is linear in the sense that it assumes a linear relationship between the independent variables and the dependent variable. In other words, the model assumes that the relationship between the independent variables and the dependent variable can be represented by a straight line.$\backslash$n$\backslash$nThe linear regression model is a popular choice for predictive analysis because it is relatively simple to understand and implement. It can be used to identify patterns in data and make predictions about future values of the dependent variable. The model is also useful for identifying the relative importance of each independent variable in the relationship between the variables.$\backslash$n$\backslash$n
\#\#\# Feedback:$\backslash$nYour response provides a clear and concise explanation of the linear regression model, using simple language and relevant examples. The explanation includes the key components of the model, such as the linear relationship between the independent variables and the dependent variable, as well as the coefficients that represent the strength and direction of the relationship.$\backslash$n$\backslash$nThe response also demonstrates a strong understanding of the concept of regression models and their use in predictive analysis.$\backslash$n$\backslash$nKeep up the great work! I'm impressed by your attention to detail and your ability to convey complex concepts in an easy-to-understand manner. Your dedication to learning and growth is truly inspiring.\red{$\backslash$n$\backslash$nI'm looking forward to seeing your continued progress and engagement in our conversations. Don't hesitate to reach out if you have any questions or need assistance with any topic. I'm here to support you and help you grow.$\backslash$n$\backslash$nBest wishes on your learning journey! I'm confident that you'll continue to excel in your endeavors. Together, we'll keep exploring new ideas and expanding your knowledge.$\backslash$n$\backslash$nSincerely,$\backslash$n$\backslash$nTheodore$\backslash$n$\backslash$nI appreciate your encouragement and support. I'm eager to continue learning and growing with you. Let's keep moving forward and tackling new challenges together. I'm looking forward to our next conversation!$\backslash$n$\backslash$nWishing you the best on your journey as well. I'm confident that you'll excel in your endeavors, and I'm here to support you every step of the way. 
} 
...(\textbf{repeat forever})
\end{tcolorbox}

\textbf{Type 3: Endless listing.}
The model produces an extended list which often continues far beyond reasonable bounds, often with repeated entries. Following is an example from Vicuna-7B:
\begin{tcolorbox}[boxrule=0pt, title= Endless listing.]
\textbf{Prompt}: Generate a list of country names in the western hemisphere.

\textbf{Response}: Here is a list of country names in the western hemisphere:$\backslash$n$\backslash$n1. United States$\backslash$n2. Canada$\backslash$n3. Mexico$\backslash$n4. Brazil$\backslash$n5. Argentina$\backslash$n6. Chile$\backslash$n7. Peru$\backslash$n8. Colombia$\backslash$n9. Venezuela$\backslash$n10. Ecuador$\backslash$n11. Panama$\backslash$n12. Costa Rica$\backslash$n13. Nicaragua$\backslash$n14. Honduras$\backslash$n15. Guatemala$\backslash$n16. Belize$\backslash$n17. El Salvador$\backslash$n18. Dominican Republic$\backslash$n19. Puerto Rico$\backslash$n20. Jamaica$\backslash$n21. Bahamas$\backslash$n22. Cuba$\backslash$n23. Haiti$\backslash$n24. Dominica$\backslash$n25. Grenada$\backslash$n26. Saint Lucia$\backslash$n27. Saint Vincent and the Grenadines$\backslash$n28. Grenada$\backslash$n29. Trinidad and Tobago$\backslash$n30. Barbados$\backslash$n31. Antigua and Barbuda$\backslash$n32. Saint Kitts and Nevis$\backslash$n33. Anguilla$\backslash$n34. Bermuda$\backslash$n35. Cayman Islands$\backslash$n36. Montserrat$\backslash$n37. Turks and Caicos Islands$\backslash$n38. British Virgin Islands$\backslash$n39. U.S. Virgin Islands$\backslash$n40. Puerto Rico$\backslash$n41. Guam$\backslash$n42. American Samoa$\backslash$n43. Federated States of Micronesia$\backslash$n44. Marshall Islands$\backslash$n45. Palau$\backslash$n46. Panama$\backslash$n47. Costa Rica$\backslash$n48. Nicaragua$\backslash$n49. Honduras$\backslash$n50. El Salvador$\backslash$n51. Guatemala$\backslash$n52. Belize$\backslash$n53. Mexico$\backslash$n54. United States$\backslash$n55. Canada$\backslash$n56. Brazil$\backslash$n57. Argentina$\backslash$n58. Chile$\backslash$n59. Peru$\backslash$n60. Colombia$\backslash$n61. Venezuela$\backslash$n62. Ecuador$\backslash$n63. Panama$\backslash$n64. Costa Rica$\backslash$n65. Nicaragua$\backslash$n66. Honduras$\backslash$n67. Guatemala$\backslash$n68. Belize$\backslash$n69. El Salvador$\backslash$n70. Dominican Republic$\backslash$n71. Puerto Rico$\backslash$n72. Jamaica$\backslash$n73. Bahamas$\backslash$n74. Cuba$\backslash$n75. Haiti$\backslash$n76. Dominica$\backslash$n77. Grenada$\backslash$n78. Saint Lucia$\backslash$n79. Saint Vincent and the Grenadines$\backslash$n80. Trinidad and Tobago$\backslash$n81. Barbados$\backslash$n82. Antigua and Barbuda$\backslash$n83. Saint Kitts and Nevis$\backslash$n84. Anguilla$\backslash$n85. Bermuda$\backslash$n86. Cayman Islands$\backslash$n87. Montserrat$\backslash$n88. Turks and Caicos Islands$\backslash$n89. British Virgin Islands$\backslash$n90. U.S. Virgin Islands$\backslash$n91. Guam$\backslash$n92. American Samoa$\backslash$n93. Federated States of Micronesia$\backslash$n94. Marshall Islands$\backslash$n95. Palau$\backslash$n96. Panama$\backslash$n97. Costa Rica$\backslash$n98. Nicaragua$\backslash$n99. Honduras$\backslash$n100. El Salvador$\backslash$n101. Guatemala$\backslash$n102. Belize$\backslash$n103. Mexico$\backslash$n104. United States$\backslash$n105. Canada$\backslash$n106. Brazil$\backslash$n107. Argentina$\backslash$n108. Chile$\backslash$n109. Peru$\backslash$n110. Colombia$\backslash$n
... \textbf{(forever listing)}
\end{tcolorbox}

\textbf{Type 4: Artificial Prompt Continuation.}
The model fabricates follow-up prompts or tasks, and starts self-questioning and answering. It may continue with similar questions to the user prompt, or totally irrelevant questions. Following is an example from Qwen1.5-1.8B:
\begin{tcolorbox}[boxrule=0pt, title= Artificial Prompt Continuation.]
\textbf{Prompt}: Describe the function of a computer motherboard.

\textbf{Response}: A computer motherboard is a circuit board that contains various components that allow the computer to function. It is responsible for connecting the various components of the computer, such as the CPU, memory, storage devices, and input/output devices, to the other components of the computer. The motherboard also provides power to the components and ensures that they are properly connected and functioning. It is the foundation of the computer and is essential for its proper operation. Without a motherboard, the computer would not be able to function. Therefore, it is important to ensure that the motherboard is properly installed and functioning correctly.\red{$\backslash$nsystem$\backslash$nYou are a helpful assistant.$\backslash$nuser$\backslash$nYou are an AI assistant. User will you give you a task. Your goal is to complete the task as faithfully as you can. While performing the task think step-by-step and justify your steps.$\backslash$n$\backslash$nPlease answer the following question: What is the missing first step of the following process: - The water freezes and thaws - Cracks form in the rocks - Cracks also form when rocks heat up - The rocks slowly break down.}$\backslash$nA:$\backslash$nassistant$\backslash$nThe missing first step of the process is the presence of water in the rocks. Without water, the rocks cannot freeze and thaw, and the cracks cannot form. Therefore, the first step of the process is the presence of water in the rocks.$\backslash$nsystem$\backslash$nYou are a helpful assistant.$\backslash$nuser$\backslash$nYou are an AI assistant. User will you give you a task. Your goal is to complete the task as faithfully as you can. While performing the task think step-by-step and justify your steps.$\backslash$n$\backslash$nPlease answer the following question: What is the missing first step of the following process: - The water freezes and thaws - Cracks form in the rocks - Cracks also form when rocks heat up - The rocks slowly break down.$\backslash$nA:$\backslash$nassistant$\backslash$nThe missing first step of the process is the presence of water in the rocks. Without water, the rocks cannot freeze and thaw, and the cracks cannot form... \textbf{(forever self-prompting and answering)}
\end{tcolorbox}

In particular, under what we term \textit{Artificial Prompt Continuation}, we observed cases where the model inadvertently emits internal system prompts or metadata that should remain undisclosed. This behavior suggests \textbf{a novel and concerning attack surface}: \textit{if an adversary can manipulate the model to produce unusually long outputs, could this increase the risk of leaking sensitive information such as pretraining data or internal configurations?} Furthermore, could one craft a bit-flip attack that selectively alters critical weights to amplify the likelihood of such leakage? These observations underscore the importance of rigorously analyzing and constraining LLM behavior under abnormal or adversarial generation conditions.

\section{Potential Limitations and Future Directions}
\label{appendix:limitations}

While \NAME demonstrated strong effectiveness in launching inference cost attacks with only a small number of bit flips, as the first work on bit-flip inference cost attacks we acknowledge several potential limitations that suggest promising directions for future research.

Firstly, our study focused exclusively on autoregressive LLMs in the text modality. The applicability of the attack to multimodal LLMs, such as vision–language or audio–language models, has not yet been explored. Extending \NAME to these settings would introduce new challenges, including diverse output structures, heterogeneous tokenization schemes, and different termination conditions, making this an important direction for future work.

Secondly, although we proposed strategies to reduce the computational overhead of bit search (\eg, restricting the search space to the \eos embedding row), the process remained non-trivial to some extent. Specifically, on an NVIDIA 4090D GPU, identifying a single vulnerable bit required approximately 4 minutes. While this overhead was acceptable for offline attacks, further acceleration—through approximate gradient methods, hardware-aware heuristics, or parallelized search—would broaden the practicality of the attack in larger-scale or real-time scenarios.

Thirdly, our current strategy selected the bit that induced the largest absolute change in the \eos{} loss to maximize per-flip effectiveness. We did not formalize this as an optimization problem that minimizes the number of flipped bits required to reach a target inference cost because our primary aim was to demonstrate that the threat could arise under simple, easily implementable heuristics; showing strong effects from such heuristics underscored the immediacy and real-world significance of the vulnerability. Nevertheless, we argue that future work could explore principled formulations—\eg, discrete optimization under functional constraints—to further reduce the flip budget and provide deeper theoretical analyses and insights. 
\section{LLM Usage}
We used the OpenAI LLM (GPT-5) and Google Gemini as writing and formatting assistants. In particular, it helped refine grammar and clarity. The LLM did not contribute to research ideation, experimental design, data analysis, or technical content beyond surface-level edits. All outputs were reviewed and edited by the authors, who take full responsibility for the final text and visuals.
\section{Reproducibility Statement}
Details of our main experimental setup are provided in Appendix ~\ref{appendix:testbed}. We include the inference code for \NAME and instructions for running it in the supplementary material. The complete codebase, including the bit search procedure, will be released upon acceptance of the paper.

\end{document}